\documentclass[a4paper,11pt]{article}
\usepackage{pos}
\usepackage{hyperref}
        \hypersetup{
           breaklinks=true,   % splits links across lines
           colorlinks=true,   % displays links as colored text instead of blocks
           pdfusetitle=true,  % \title and \author values into pdf metadata
        }
\usepackage[sort&compress,numbers]{natbib}
\title{ Quantum technologies for fundamental (HE) physics}
\author*[a,b]{Diego Blas}

\affiliation[a]{Grup de Física Teòrica, Departament de Física, Universitat Autònoma de Barcelona,
08193 Bellaterra (Barcelona), Spain }

\affiliation[b]{Institut de Fisica d’Altes Energies (IFAE), The Barcelona Institute of Science and Technology, Campus UAB, 08193 Bellaterra (Barcelona), Spain}

\emailAdd{dblas@ifae.es}
\abstract{In this brief contribution I will
highlight some directions where the developments in the frontier of (quantum) metrology may be key for fundamental high energy physics (HEP). I will focus on the detection of dark matter and gravitational waves, and introduce ideas from atomic clocks and magnetometers, large atomic interferometers and detection of small fields in electromagnetic cavities. Far from being comprehensive, this contribution is an invitation to everyone in the HEP and quantum technologies communities to explore this fascinating topic.}

\FullConference{The European Physical Society Conference on High Energy Physics (EPS-HEP2023)\\
 21-25 August 2023\\
Hamburg, Germany\\}

\begin{document}
\begin{flushright}
AION-REPORT/2023-12
\end{flushright}
\maketitle

\section{Introduction}\label{sec:intro}

The field of quantum technologies for fundamental physics is going through a revolution. The reasons are multiple. First, the progress at all levels in the realm of quantum technologies motivates an intense exploration of how the high precision and control achieved can be used to detect the subtle traces of new particles or other new phenomena.  Also, different more classical approaches to the detection of dark matter or  gravitational waves are exhibiting their limits, and quantum technologies offer new exciting possibilities to extend them.
Examples of this excitement are: the CERN Quantum Technology Initiative (CERN QTI, \href{https://quantum.cern/}{https://quantum.cern/}), the Physics Beyond Colliders Study Group (\href{https://pbc.web.cern.ch/}{https://pbc.web.cern.ch/}), 
the Task Force 5 (Quantum and Emerging Technologies) of the ECFA Detector R\&D Roadmap Symposium (\href{https://indico.cern.ch/event/999818/}{https://indico.cern.ch/event/999818/}), the Fermilab Quantum Institute (\href{https://quantum.fnal.gov}{https://quantum.fnal.gov}), the UKRI Quantum Technologies for Fundamental Physics (\href{https://uknqt.ukri.org/our-programme/qtfp/}{https://uknqt.ukri.org/our-programme/qtfp/}), the Fundamental Physics Program Analysis Group (FunPAG) of NASA (\href{https://www.jpl.nasa.gov/go/funpag}{https://www.jpl.nasa.gov/go/funpag}) or the reviews \cite{Ahmed:2018oog,Safronova:2017xyt,Bass:2023hoi}\footnote{If the reader misses other interesting initiatives/references, please contact me to add them.}.

In this contribution I shall focus on the metrology and/or quantum sensing part of this effort when applied to the search of new fundamental backgrounds. The first great advantage of this approach as compared to other detectors (e.g. those that need some minimal momentum transferred to generate a signal) is that quantum mechanically any scattering event leaves an imprint in the final states (e.g. in their phases), even when the momentum transferred vanishes \cite{sakurai2017modern}. In this sense, large fluxes of low-energy particles are an ideal target for these technologies. Furthermore, some of the intrinsic quantum phenomena (such as entanglement) are fragile, and subject to work as sensors of any interacting background. The combination of this fact with the possible presence of coherent   effects may also produce very sensitive detectors. Finally, quantum noise is already playing a role in several measurements, e.g. in ground-base interferometers \cite{Evans:2013aua}. The ability to manipulate our sensors may beat this quantum noise in some situations, e.g. \cite{Dwyer:2022vbh}, and understanding the limitations of how much the quantum noise can be reduced is essential for future detectors. 

Another important aspect of the program is that its boundaries are far from fixed. The fast development of our ability to understand and manipulate quantum phenomena is spurring the imagination of several groups that try to leverage it towards new detection possibilities. As an example of this development, one can consider atomic clocks: the fractional frequency uncertainty has been consistently improving since 1960s in the case of microwave clocks, and very dramatically since 1990s for optical clocks 
\cite{Optical}. When considering the growing investment into the field (compare e.g. this study\footnote{Another source of information of the global effort in quantum technologies is \href{https://thequantuminsider.com/2023/02/17/quantum-technology-2022-investment-update-key-trends-and-players/}{https://thequantuminsider.com/2023/02/17/quantum-technology-2022-investment-update-key-trends-and-players/}.} for 2023: \href{https://qureca.com/overview-of-quantum-initiatives-worldwide-2023/}{https://qureca.com/overview-of-quantum-initiatives-worldwide-2023/} with the similar one from 2020: \href{https://qureca.com/es/the-quantum-ecosystem-and-its-future-workforce-2/}{https://qureca.com/es/the-quantum-ecosystem-and-its-future-workforce-2/}), the possibilities for the future seem extraordinary. 

Before moving to the main body of this contribution I want to make a couple of remarks. First, I want to acknowledge that quantum technologies have been used in HEP detectors for many years. The main difference of more recent effort is that the latter put the \emph{quantum technology at the center of the detection strategy}, which typically also means focusing on table-top set-ups (at least as a starting point). Second, a sub-field which I find very promising, but that I will touch only very briefly, is the use of quantum sensors to detect ``machine-made'' backgrounds (e.g. by placing them close to reactors or beam-dumped experiments). The small size of these devices and their sensitivity to small momentum transfer may be key in searches for new physics in these set-ups.

\section{Fundamental cosmological backgrounds}\label{sec:back}

Our Universe has been rather generous in the fundamental cosmological (and astrophysical) backgrounds that arrive to Earth. One of the best known examples is the cosmic microwave background (CMB) generated when recombination happened and the Universe became transparent \cite{Durrer:2020fza}. This signal (today detected in the microwave band) carries very precise information from 13.7$\times 10^9$ years ago, but also from the intervening structure that it found as  travelling towards us.
%The detection of this background was awarded the Nobel prize in Physics of 1978 and the discover of its primary anisotropy the Nobel prize in Physics of 2006.

How do we access more primordial information or  from places that have obscured the CMB signal or that do not produce detectable photons? A possible answer is by using \emph{neutrinos}, that can have very long mean-free paths and for which the Universe was almost transparent at recombination. The equivalent of the CMB in this case is the cosmic neutrino background (CNB): as the Universe's temperature dropped below few MeV (which happened much earlier than recombination), the neutrinos that participated in nuclear reactions saw an almost transparent universe, and propagate  from this moment almost unimpeded \cite{Lesgourgues:2013sjj}. The searches for the CNB have so far proved fruitless, in part due to the small typical momentum of its constituents, which means a tiny available energy to generate recoils  and very small cross-section of the CNB with our detectors. One of the most promising directions is PTOLEMY  \cite{PTOLEMY:2019hkd}, though it is not yet clear when the CNB will be found (see for instance \cite{Cheipesh:2021fmg,Bauer:2022lri,Arvanitaki:2023fij}). The quantum frontier may certainly be explored to this aim and to answer other questions about neutrinos: their mass, their nature (Dirac or Majorana), more insights on their family structure or possible new interactions with standard model fields (photons, in particular). For this direction, one could not only use cosmological neutrinos, but also machine-made fluxes. As described above, I will not discuss this fascinating possibility in any detail in this contribution. 

A more uncertain, yet existing, background is that of gravitational waves (GWs). GWs are in principle produced by every energetic process in the Universe. This universality, and the fact that they travel without basically interacting through cosmological scales, makes of them ideal messengers of  some of the most extraordinary events taking place in the cosmos. For instance, they are copiously generated in the dynamics of many scenarios of the primordial Universe (inflation in particular) \cite{Caprini:2018mtu} or by the merger of black holes at more recent times, that do not leave any other observable signal. Their first direct detection took place in 2015, and it corresponded to GWs in a  window close to  100 Hz and emitted by a merger of black holes \cite{LIGOScientific:2016aoc}. Since then, many other detections have followed in  the 10 Hz to 10 kHz band. In 2023, a new signal was confirmed at nHz \cite{EPTA:2023fyk,Reardon:2023gzh,NANOGrav:2023gor,Xu:2023wog}. There is a global effort to detect GWs at other frequencies, with CMB studies \cite{Planck:2018jri} and the ESA lead mission LISA \cite{LISA:2017pwj} as spearheads. Quantum technologies may impact this search in several ways. Besides increasing the sensitivity of current detectors, they can play a pivotal role to access \emph{higher frequencies}, which may have a direct impact in laboratory set-ups. I will discuss this further in the next section. A better coverage of the spectrum of GWs will have a very strong impact in our understanding of Nature, which motivates the widest possible search.

The final possible fundamental background I will discuss is that coming from the dark sector. Our current understanding of the Universe requires the existence of a dark matter (DM) component, five times more abundant than ordinary matter, and a dark energy component controlling the current accelerated expansion observed at cosmological distances \cite{ParticleDataGroup:2020ssz}. Very little more is known about them, besides their expected energy density at different places, including the Solar system \cite{Nesti:2013uwa}. Properties such as the mass, spin, interactions with standard model fields, state (classical field, particle, compact object...)... of the particles that may be responsible of these `dark' sectors are rather unconstrained. Coming back to the logic of Sec.~\ref{sec:intro}, the possibility to control very fragile systems opens a battery of new techniques to directly access the dark sector. In the next section I will only discuss the influence of some DM models in selected set-ups.  It would be very useful  to extend this to all possible models (of DM, but also of dark energy) and set-ups.

\section{Deriving the influence of fundamental backgrounds in precision measurements}

%The previous sections where rather vague and generic. 
When one wants to take the next step in the program that I am describing, a \emph{model agnostic} approach
is possible, based on ideas from 
%to select a model and study all its possible effects in a (quantum) device. The lesson from recent years in phenomenology of particle physics and  theories of gravitation shows that another more fruitful approach is possible. Indeed, one can be 
%and address the problem from an 
\emph{effective field theory}. The idea can be summarized in the following steps: 
\emph{i)} One decides the fundamental properties of the field of interest (e.g. spin or mass).
 \emph{ii)} One decides the symmetries that govern the interactions.
\emph{iii)} One writes \emph{all} possible interactions of these new fields with ordinary matter, and classifies them according to their relevance. Here want to understand the coupling to the degrees of freedom in laboratory set-ups, which are normally related to atomic physics and light.
\emph{iv)} One derives the limit of interest for the  laboratory set-ups. Typically, this means the non-relativistic approximation for matter.
\emph{v)} One studies the consequences of these operators (``interaction Hamiltonians'') one by one, and for the different possible states of the background (considered e.g. as a classical field, collection of particles, etc.). Possible effects may be those of new absorption channels, scattering processes, decays, etc.
\emph{vi)} One finally connects the conclusions to the possible models including them. 
%\begin{enumerate}
%\itemsep -3pt
 %   \item One decides the fundamental properties of the field of interest (e.g. spin or mass).
  %  \item One decides the symmetries that govern the interactions.
   % \item One writes \emph{all} possible interactions of these new fields with ordinary matter, and classify them according to their relevance. This last step is slightly different from what is done in e.g. collider searchers. Here we only want to understand the coupling to the degrees of freedom of interest in laboratory set-ups, which are normally related to atomic physics and light.
    %\item One derives the limit of interest for the  laboratory set-ups. Typically, this means the non-relativistic approximation for matter.
    %\item One studies the consequences of these operators (``interaction Hamiltonians'') one by one, and for the different possible states of the background (e.g. considered as a classical field, collection of particles, etc.). Possible effects may be those of new absorption channels, scattering processes, decays, etc.
    %\item One finally connects the conclusions to the possible models including them. 
%\end{enumerate}

I believe that this program can be addressed in a rather generic way, and generate a useful dictionary to catalyze the communication between the HEP and the quantum sensing communities. 
I will now use the interaction of light with matter as an example of how this could work. Let us imagine that we want to detect photons without knowing their properties, beyond the fact that they are spin-1 particles (not even their mass). One can postulate that they  couple to electrons through an associated operator. The set of allowed operators in $A^\mu$, in increasing dimension, starts as
\begin{equation}
\label{operators}
   q_{1i} A^\mu \bar \psi_e O^i_\mu \psi_e,~~~~~  q_{2i}\frac{  \partial^\nu A^\mu}{\Lambda} \bar \psi_e O^i_{\mu\nu} \psi_e,~~~~ q_{3i}\frac{  A^\nu A^\mu}{\Lambda} \bar \psi_e \tilde O^i_{\mu\nu} \psi_e, ...
\end{equation}
where $O^i_\mu$,   $O^i_{\mu\nu}$ and $\tilde O^i_{\mu\nu}$ are operators acting on the spinor $\psi_e$ and associated to   a unitary theory. $\Lambda$ is a scale that controls the relevance of the second and third operators with respect to the first one (we have assumed that  all $q_{ni}$ have the same dimensionality, as do $O_\mu$, $O_{\mu\nu}$ and $\tilde O^i_{\mu\nu}$). These coefficients are a priori arbitrary and their effects may be studied independently. If one  focuses on electrodynamics, then it is the term with $O^i_\mu=\gamma_\mu$ which is relevant.  This analogy is very important, because some of the effects we may be looking for have already been considered as possible backgrounds or foregrounds for the devices. For instance, in \cite{Gibble} the effect of a background gas collisions is studied for atomic clocks. The last operator in \eqref{operators} may be constrained with this observation. For quark couplings, one needs to connect them to nuclei through some standard (quite complex) techniques, that are eventually summarize in a collection of form factors, e.g. \cite{Cerdeno:2010jj}.

The forth step  would be to reduce these operators to those relevant in the field of interest. For instance, in atomic physics our electrons will be (most of the times) non-relativistic, so we can simply evaluate the biliniars of the form $\bar \psi_e O^i_\mu \psi_e$ for arbitrary spinors, and take the non-relativistic limit, e.g. \cite{sakurai1967advanced}. The final result will be a set of non-relativistic operators, depending on the mass($m$), momentum ($\vec p$), position ($\vec x$) or spin ($\vec S$) of the electron (further contributions may appear for non-fundamental particles), 
\begin{equation}
\label{operators2}
   \tilde q_{1} A^i p^i_e,~~~~~  \tilde q_{2}\frac{ \epsilon_{ijk}\partial_j A_k}{\Lambda} S^i_e,~~~~ \tilde q_{3}\frac{  A_i A_j}{\Lambda} x^i_e  x^j_e, ...
\end{equation}
For the fifth step, one also needs to decide the state of the new field ($A^\mu$ in the previous example). It can be a massive or massless field, which may imply that it may also be described in a non-relativistic regime. In this case, one can proceed as for the fields of the detector. More relevant are the properties as a function of time, and in terms of the occupation number of its constituents. %The dark matter example is very rich.  
For instance, the field $A^\mu$ may be in  a configuration with large occupation numbers and random phases per phase space element, and a classical description may be convenient. The new field  $A^\mu$  may also interact as \emph{individual particles}, and  treated as a quantum field. In this case, each of these particles may go through a region of the device, which one can quantify by understanding the size of the corresponding wave-packet, and the average effect of the flux of incoming particles.  This requires to study in detail how the whole wave function of the elements of the device evolve with time. An example that may be illuminating is the work for atomic clocks and dark matter particles in \cite{Wolf:2018xlz,Alonso:2018dxy}. With these operators, we can proceed and study their consequences (with our quantum technologies colleagues). The final step reinterprets these effects in terms of the original proposals. 

%In some cases (as for most situations for gravitational waves and ultra-light dark matter), the occupation number of the new fields is so large that a the previous description is not the most adequate. Assuming that these states have random phases, one would conisder the fields in \eqref{operators2} as classical, and proceed to search for their effects. 

For this fifth step, one needs to use several properties of the backgrounds. For, instance for dark matter we expect an energy density of $\rho\approx 0.3\,$GeV/cm${}^3$ and an average velocity of $\bar v\approx 10^{-3}$ due to the motion of the Sun in an otherwise basically isotropic background of DM, which can produce effects of the from
$\alpha S_e^i m v_{\rm DM}^i$ If this is the case, the fact that the average velocity does not cancel, will generate a non-vanishing coupling that may be confused with an anomalous magnetic field. However, there are handles to break this degeneracy: the coupling to different species of standard fields, or the fact that the effect will be modulated yearly and daily as we travel through the DM galactic halo. Note also that even in the case of a non-coherent total effect, this term will act as a \textit{noise source} for the determination of spin measurements, eventually acting as a source of uncertainty that one can try to search for. 

The previous steps can be formalized and organized to  fit a single review where all (or almost all) relevant effects may be summarized.  This sort of \emph{dictionary}  would be very useful for a fruitful dialogue with the community of quantum technologies and I hope that we are able to do this in the upcoming years.

\section{Three (biased) examples}

In this section, I will briefly comment on three examples where the new frontiers in (quantum) metrology  are playing a relevant role for the program I described above. 

\subsection{Dark matter and cosmic neutrinos with atomic clocks and co-magnetometers}

It is well-known that light dark matter may not have enough momentum to generate recoil events in traditional dark matter detectors. More concretely, for dark matter scattering with a nucleus, the recoil energy is at most $E_R^{\max } \sim\left(\frac{m_\chi}{\mathrm{GeV}}\right) \mathrm{keV} $, which may be below threshold to be observed in traditional searches \cite{Lisanti:2016jxe}. The quest to observe light dark matter requires new ideas \cite{Battaglieri:2017aum}. The reason why quantum detectors may yield a new handle in this problem is that \emph{even at vanishing momentum transfer} the scattering of particles has an impact on the phase of the scattering process. Hence, there is no real threshold in these experiments, as long as we are able to detect the effect of the scattering in the phase of the scattered particles\footnote{This is not exactly true, as the phase can not be determined with arbitrary precision in a system with a finite number of degrees of freedom \cite{Carruthers:1968my}.
See e.g. \cite{Beckey:2023shi} for a recent discussion of quantum measurements in fundamental physics.}. This was studied in detail in \cite{Wolf:2018xlz,Alonso:2018dxy} (see also \cite{Du:2022ceh}). In these works, we studied how the transition probability is modified in an interferometry process by the scattering of DM particles of very light mass. The latter is the basis for atomic clocks and atomic magnetometers, and our results show how these set-ups may yield spectacular constraints for many models of particle DM\footnote{Atomic clocks and magnetometers have been proposed to look for ultra-light dark matter, where the DM component behaves as a classical field \cite{Arvanitaki:2014faa,Bloch:2019lcy}. We still work in the particle regime, and it is the collection of many scattering events which generates the signal.}. I will only briefly mention the case of co-magnetometers, and refer the reader to the previous references for more details. Atomic magnetometers are based on a sample of $N\sim 10^{21}$ polarized atoms, whose spin interacts with an external magnetic field through an operator
$%\begin{equation}
    H_{\rm int}=-\gamma \vec B\cdot \vec \lambda, \label{H}
$%\end{equation}
where $\lambda$ represents the spin of the atom, and $\gamma$ is the effective coupling. As a result, for a polarized sample where $\vec\lambda$ is not aligned with $\vec B$, $\vec \lambda$ of the atoms will precess with a frequency $\omega\equiv \gamma B$, which is connected to the difference in the energy levels generated by \eqref{H}. If a flux of DM goes through the sample and scatters through it, the forward scattering limit can be interpreted as a modification of the effect of the free Hamiltonian, and one can show that the effective Larmor frequency is now 
\begin{equation}
    \omega_{\rm DM} \equiv \gamma \beta=\gamma\left(B+\frac{2 \pi n_\chi}{m_\chi \gamma}\left(\bar{f}(0)_1-\bar{f}(0)_2\right)\right),
\end{equation}
where $f(0)_i$ is the matrix element corresponding to forward scattering of the different spin states with $i$. Here $n_\chi$ is the DM number density and $m_\chi$ the DM mass and the bar indicates the average over the scattering events during the experiment. This last concept is quite different as compared to other searches: since we are sensitive to the changes in the phase, it is important that the scattering events do not average to 0\footnote{As mentioned above, a certain level of induced ``noise'' is always expected. Furthermore, other effects, such as the lost of coherence, are directly affected by the total number of scattering events.} , which may happen if the interaction depends on arbitrary parameters such as the DM spin. As far as I know, this approach has not been fully exploited to search for DM, and I invite the reader to find more details in  \cite{Wolf:2018xlz,Alonso:2018dxy,Du:2022ceh}. Similarly, it seems a good possibility to find new ways to explore the CNB. Our results of \cite{Alonso:2018dxy} implied that we could win several orders of magnitude if the CNB is polarized or has some non-trivial chemical potential. Even in these non-standard cases, we are very far from claiming that this is a promising direction. 

Before moving to the next example, I want to mention that these ideas could also be used to detect large backgrounds/fluxes which may be produced in reactors or accelerators (e.g. at beam-dumped configurations, such as those described in \cite{BDX:2017jub}), and go through the quantum devices without the need to deposit their energy. I am not aware of any publication in this direction. 

\subsection{Large atomic interferometers}

The use of large atomic interferometers (AI) to detect dark matter or gravitational waves has received growing attention in the last years, e.g. \cite{Proceedings:2023mkp,Buchmueller:2023nll} (see also \cite{Dimopoulos:2008sv,Arvanitaki:2016fyj,Dimopoulos:2007cj,Canuel:2019abg,Badurina:2019hst}).
The idea of these techniques can be summarized as follows: a sample of atoms, which can be  in states $|e\rangle$ and  $|g\rangle$ with energy difference $\omega$, are launched  in the $|g\rangle$ state and evolve in free fall, except at some times when a laser generates $\pi/2$ or $\pi$ pulses. The first of these pulses (at $t=0$) split the beam into $|g\rangle$ and $|e\rangle$ atoms and imprint a momentum kink to these last ones. The $\pi$ pulse (at $t=T$), when performed carefully, turns the $|e\rangle$  portion of the sample into $|g\rangle$, and viceversa, while generating  a negative  momentum kick to the $|e\rangle$ states, and positive to the $|g\rangle$ states. Finally, another $\pi/2$ pulse is performed at $t=2T$ (see  \cite{Proceedings:2023mkp,Buchmueller:2023nll,Dimopoulos:2008sv,Arvanitaki:2016fyj,Dimopoulos:2007cj,Canuel:2019abg,Badurina:2019hst} for more details). The final result of all these processes, if unperturbed, generates a wavefunction with different phases for the two samples of atoms
\begin{equation}
    \left.|\psi\rangle \sim \frac{1}{\sqrt{2}}\left(\left(1+e^{-i \Phi_{\mathrm{TOT}}}\right)|g\rangle+\left(1-e^{-i \Phi_{\mathrm{TOT}}}\right)|e\rangle\right)\right).
\end{equation}
In the previous expression $\Phi_{\mathrm{TOT}}\sim \omega T$, where $T$ is defined above. As a result, the frequency $\omega$ is imprinted in the final state, and by measuring the population in each state, we can be sensitive to the energy difference and $T$. When we consider that the previous processes happen in the presence of a DM background, the same physics considered in the previous section does in principle modify the energy split between the two states if the DM particle distinguishes between them as it scatters. For instance,  \eqref{H} would generate an extra shift in the energy difference of the states of different spin. Other DM candidates may simply change the fundamental constants, and hence the energy levels. For instance,  a coupling $\phi m_e \bar \psi_e
\psi_e$ (quite standard for ultra-light dark matter \cite{Buchmueller:2023nll}) would change the mass of the electron if the field $\phi$ has some time or space-dependent properties (e.g. if it behaves as a non-trivial classical field). As should be apparent from the the previous section, the scattering of DM of higher masses can also generate interesting phenomena in AI \cite{Du:2022ceh}. I believe that there is still plenty of exploration space in this direction. For instance, AIs are also  fantastic accelerometers, and this may yield a new handle to discover light dark matter \cite{Domcke:2017aqj}.

Another very interesting use of AIs is as explorers of GWs. This is simple to understand from the fact that the the light travel time $L$ in  will be modified in the presence of GWs, impacting the previous processes. The relevance of this possibility is enhanced by the fact that the set-ups that have been suggested for the following years will be sensitive to GWs in a band very hard to cover by other methods (a bit below Hz, the mid-band between space-based and ground-based detectors) \cite{Proceedings:2023mkp,Badurina:2021rgt}.

To the interested reader, here is a (probably incomplete) selection of proposals:  the MAGIS-100 detector 
\cite{MAGIS-100:2021etm} (currently developing a 100-m vertical set-up); the MIGA project \cite{Canuel:2017rrp} (currently developing a horizontal set-ups of $\sim 200$m); the ELGAR initiative 
\cite{Canuel:2019abg}; the ZAIGA proposal \cite{Zhan:2019quq} (aiming at a 300-m vertical set-up); the AION collaboration \cite{Badurina:2019hst} (currently building the 10-m vertical set-up); and the futuristic AEDGE \cite{AEDGE:2019nxb}, proposing atomic interferometry in space. 
A lot of new collaborations are possible, in particular regarding where to install the future set-ups \cite{Buchmueller:2023nll,Arduini:2023wce}.

\subsection{ GWs in (superconducting radio-frequency) cavities}

I will conclude with a brief discussion of another idea where cutting-edge (quantum) metrology may play a very important role for fundamental physics. As I discussed in Sec.~\ref{sec:back}, the study of the widest possible spectrum of GWs is a very important challenge of modern physics. By accessing frequencies above 10 kHz (those scrutinized with ground-based detectors -- the LIGO-Virgo-KAGRA collabororation), one explores a  plethora of dark matter and early universe models very hard to probe otherwise \cite{Aggarwal:2020olq}. Since signals in the range MHz to GHz or even higher are conceivable, this implies a direct connection with the time and length scales of laboratory set-ups, and hence an exciting possibility of testing high frequency gravitational waves (HFGWs) with quantum sensors. 

Form the equivalence principle, we know that any source of energy and momentum couples to gravitation. As a result, an  abundance of effects can be expected to be induced by the passage of a HFGW in a very precise device. I will only comment on one of these ideas: the effect of a HFGW as it impacts an electromagnetic cavity in the presence of background electromagnetic (EM) fields $A_b^\nu$. Two main processes may occur: the first one is that as the GW interacts with the background EM field, it generates an effective current $j_\mu$ linear in $h^{\alpha\beta}$ and $A_b^\nu)$ that sources a new ``signal'' EM field. This operator $A^\mu_s j_\mu$ is easy to understand as the energy and momentum that GWs couple to are quadratic in $A^\mu$. Quite relevantly, this coupling is similar to the one natural for  axionic dark matter ($a$), $a\, \vec E\cdot \vec B$ \cite{Raffelt:1987im}, which means that a whole battery of techniques searching for these fields is ready to be used to search for GWs. Furthermore, the GW shakes the walls of the cavity\footnote{GWs of amplitude $h$ change ``physical'' distances as $\delta L\sim h L$.}, changing its geometry and affecting the properties of the modes that may be stored in $A_b^\nu$ (which may generate mode-mixing). Due to lack of space (time in my talk), I cannot describe these set-ups in more detail. I would still convey the message that current studies show a vast parameter space that may be explored in the near future, where new surprises from Nature may be waiting \cite{Berlin:2021txa,Berlin:2023grv,Domcke:2023qle,Aggarwal:2020olq}.

\section{Summary and future directions}

Quantum sensing/devices provide new ways to detect foregrounds/backgrounds of relevance for fundamental physics
(e.g. gravitational waves, neutrinos or dark matter and dark energy). A key characteristic of why this can generate breakthroughs in the near future are the low energy/momentum transferred thresholds for quantum effects to happen,  which are ideal to sense ``substantial'' fluxes with tiny cross-sections. 
I believe that as HEP practitioners, we have a very important task in front of us to transform our models of interest into predictions for the degrees of freedom that (quantum) sensing practitioners recognize. In particular, by generating a dictionary from HEP models into interacting Hamiltonians $H = H_0 + H_{\rm int}^{\rm new\ physics}$ we may bridge an important gap and generate a fluid dialogue between communities.

In this short (and inevitably biased) contribution, I presented some examples where we started to fill this gap for searches of dark matter and GWs in co-magnetometers, atomic clocks, long-baseline interferometers and EM cavities (with a couple of comment  on  beam-dumped experiments and neutrino searches).  I am convinced that we are just starting to grasp the potential of  the quantum world for fundamental physics, and the use of ideas such as faster readouts,  single-photon detectors, larger or entangled samples, will take us towards totally unexpected consequences. I hope that this contribution  is intriguing and interesting enough to motivate future directions towards new milestones in our understanding of Nature. 

{\small \textbf{Acknowledgements.}  It is a pleasure to thank L. Badurina,  S. Ellis and C. Murgui for comments on a first draft of this contribution. I am supported  by a `Ayuda Beatriz Galindo Senior' from the Spanish `Ministerio de Universidades', grant BG20/00228. The research leading to these results has received
funding from the Spanish Ministry of Science and Innovation
(PID2020–115845GB-I00/AEI/10.13039/501100011033). IFAE is partially funded by the CERCA program of the Generalitat de Catalunya. I also acknowledge the support from the Departament de Recerca i Universitats de la Generalitat de Catalunya al Grup de Recerca i Universitats from Generalitat de Catalunya to the Grup de Recerca 00649 (Codi: 2021 SGR 00649). }

\end{document}